\documentclass[traditabstract,letter]{aa} 
\usepackage{txfonts}
\usepackage{graphicx}
\usepackage{natbib}

\newcommand{\onn}{\mbox{\object{NN\,Ser}}}
\newcommand{\onnab}{\mbox{\object{NN\,Ser\,ab}}}
\newcommand{\nnser}{NN\,Ser}
\newcommand{\nnserab}{NN\,Ser\,ab}

\newcommand{\ac}{$a_\mathrm{c}$}
\newcommand{\ad}{$a_\mathrm{d}$}
\newcommand{\ec}{$e_\mathrm{c}$}
\newcommand{\ed}{$e_\mathrm{d}$}
\newcommand{\pc}{$P_\mathrm{c}$}
\newcommand{\pd}{$P_\mathrm{d}$}
\newcommand{\omegac}{$\varpi_\mathrm{c}$}
\newcommand{\omegad}{$\varpi_\mathrm{d}$}
\newcommand{\massc}{$M_\mathrm{c}$}
\newcommand{\massd}{$M_\mathrm{d}$}
\newcommand{\rp}{$r_\mathrm{p}$}
\newcommand{\rsun}{R$_\odot$}
\newcommand{\msun}{M$_\odot$}
\newcommand{\mjup}{M$_\mathrm{Jup}$}
\newcommand{\chisq}{$\,\chi^2$}
\newcommand{\chisqr}{$\,\chi^2_\nu$}

\begin{document}

\title{Two planets orbiting the recently formed
 post-common envelope binary NN Serpentis}

\author{
K. Beuermann\inst{1}
\and F. V. Hessman\inst{1}
\and S. Dreizler\inst{1}
\and T.~R. Marsh \inst{2}
\and S.~G. Parsons \inst{2}
\and D. E. Winget\inst{3}
\and G.~F. Miller\inst{3}
\and M. R. Schreiber\inst{4}
\and W. Kley\inst{5}
\and V.~S. Dhillon \inst{6}
\and S.~P. Littlefair \inst{6} 
\and C.~M. Copperwheat \inst{2}
\and J.~J. Hermes\inst{3}
}

\institute{
Institut f\"ur Astrophysik,
Georg-August-Universit\"at, Friedrich-Hund-Platz 1, D-37077
G\"ottingen, Germany,
\and
Department of Physics, University of Warwick, Coventry, CV4\,7AL,UK
\and
Dept. of Astronomy, University of Texas at Austin,
RLM 16.236,  Austin, TX 78712, USA 
\and 
Departamento de Fisica y Astronomia, Universidad de Valparaiso,
Av. Gran Bretana 1111, Valparaiso, Chile,
\and
Institut f\"ur Astronomie \& Astrophysik, Universit\"at T\"ubingen,
Morgenstelle 10, D-72076 T\"ubingen, Germany,
\and
Department of Physics \& Astronomy, University of Sheffield, S3\,7RH, UK
}

\date{Received 26 July 2010 / accepted 6 October 2010}

\authorrunning{K. Beuermann et al.} 
\titlerunning{Two planets orbiting NN Serpentis}

\abstract { Planets orbiting post-common envelope binaries provide
 fundamental information on planet formation and evolution.  We
 searched for such planets in NN\,Ser\,ab, an eclipsing short-period
 binary that shows long-term eclipse time variations.  Using
 published, reanalysed, and new mid-eclipse times of NN\,Ser\,ab
 obtained between 1988 and 2010, we find excellent agreement with the
 light-travel-time effect produced by two additional bodies
 superposed on the linear ephemeris of the binary. Our
 multi-parameter fits accompanied by N-body simulations yield a best
 fit for the objects NN\,Ser\,(ab)c and d locked in the 2\,:\,1 mean
 motion resonance, with orbital periods
 $P_\mathrm{c}\!\simeq\!15.5$\,yrs and $P_\mathrm{d}\!\simeq\!7.7$
 yrs, masses
 $M_\mathrm{c}\,\mathrm{sin}\,i_\mathrm{\,c}\!\simeq\!6.9$\,\mjup\ and
 $M_\mathrm{d}\,\mathrm{sin}\,i_\mathrm{\,d}\simeq\!2.2$\,\mjup, and
 eccentricities $e_\mathrm{c}\!\simeq\!0$ and
 $e_\mathrm{d}\!\simeq\!0.20$. A secondary \chisq\ minimum
 corresponds to an alternative solution with a period ratio of
 5\,:\,2.  We estimate that the progenitor binary consisted of an A
 star with $\sim\!2\,M_\odot$ and the present M dwarf secondary at an
 orbital separation of $\sim\!1.5$\,AU. The survival of two planets
 through the common-envelope phase that created the present white
 dwarf requires fine tuning between the gravitational force and the
 drag force experienced by them in the expanding envelope. The
 alternative is a second-generation origin in a circumbinary disk
 created at the end of this phase. \mbox{In that case, the planets
   would be extremely young with ages not exceeding the cooling age
   of the white dwarf of $10^{\,6}$\,yrs.}  } 
{}{}{}{}


\maketitle

\section{Introduction}

\onnab\footnote{On recommendation by the Editor of A\&A, we refer
 to the system as NN\,Ser, to the binary explicitly as NN\,Ser\,ab,
 and to the objects orbiting the binary as NN\,Ser\,(ab)c and
 NN\,Ser\,(ab)d.}  is a short-period
($P_\mathrm{orb}$\,=\,$3.12$\,hr) eclipsing binary at a distance of
500 pc. The detached system contains a hot hydrogen-rich white dwarf
NN\,Ser\,a of spectral type DAO1 and an M4 dwarf star NN\,Ser\,b with
masses of 0.535\,\msun\ and 0.111\,\msun, respectively
\citep{parsonsetal10a}. With an effective temperature of 57000\,K
\citep{haefneretal04}, the white dwarf has a cooling age of only
$10^6$\,yrs \citep{wood95}. The present system resulted from a normal
binary with a period of $\sim$1 year when the more massive component
evolved to a giant and engulfed the orbit of its companion. The
subsequent common envelope (CE) phase led to the expulsion of the
envelope, laying bare the newly born white dwarf and substantially
shortening the orbital period.

Some eclipsing post-CE binaries display long-term eclipse time
variations, among them V471\,Tau \citep{kaminskietal07}, QS\,Vir and
NN\,Ser \citep[][and references there\-in]{parsonsetal10b}.  The
latter possesses deep and well-defined eclipses, which allow
measurements of the mid-eclipse times to an accuracy of 100\,ms and
better \citep{brinkworthetal06,parsonsetal10b}. The processes advanced
to explain them include the long-term angular momentum loss by
gravitational radiation and magnetic braking, possible
quasi-periodicities caused, e.g., by Applegate's (1992) mechanism, and
the strict periodicities produced by apsidal motion or the presence of
a third body in the system. Finding the correct interpretation
requires measurements of high precision and a coordinated effort over
a wide range of time scales. The existence of a third body orbiting
\nnserab\ was previously considered by \citet{qianetal09}, but the
orbital parameters suggested by them are incompatible with more recent
data \citep{parsonsetal10b}. In this Letter, we present an analysis
of the eclipse time variations of \nnserab, based on published data,
the reanalysis of published data, and new measurements obtained over
the first half of 2010.

\section{The data}

After their 1988 discovery of deep eclipses in \nnser,
\citet{haefneretal04} acquired a series of accurate mid-eclipse times
in 1989. After a hiatus of ten years, they added a potentially very
accurate trailed CCD imaging observation using the ESO VLT.  From 2002
on, the Warwick group systematically secured a total of 22 mid-eclipse
times of high precision \citep[][this
 work]{brinkworthetal06,parsonsetal10b}. \citet{parsonsetal10b} list
all published mid-eclipse times by other authors until the end of
2009. These are included in our analysis that weights them by their
statistical errors. Since the individual Warwick mid-eclipse times
between 2002 and 2009 were separated by about one year, information on
eclipse time variations on a shorter time scale is lacking.  We,
therefore, organized a collaborative effort of the G\"ottingen,
McDonald, and Warwick groups to monitor \onn\ over the first half of
2010. We used the remotely controlled MONET/North 1.2-m telescope at
McDonald Observatory via the MONET internet remote-observing
interface, the McDonald 2.1-m telescope, and the ESO 3.5-m NTT. The
MONET data were taken in white light, the McDonald data with a BG40
filter, and the NTT observations were acquired with the ULTRACAM
high-speed CCD camera equipped with Sloan filters. The mid-eclipse
times measured in Sloan u'$\!$, g'$\!$, and i'$\!$ are consistent, and
we used the g' data as the most accurate set for the present purpose.

\begin{figure}[t]
\includegraphics[bb=165 59 540 699,height=89mm,angle=-90,clip]{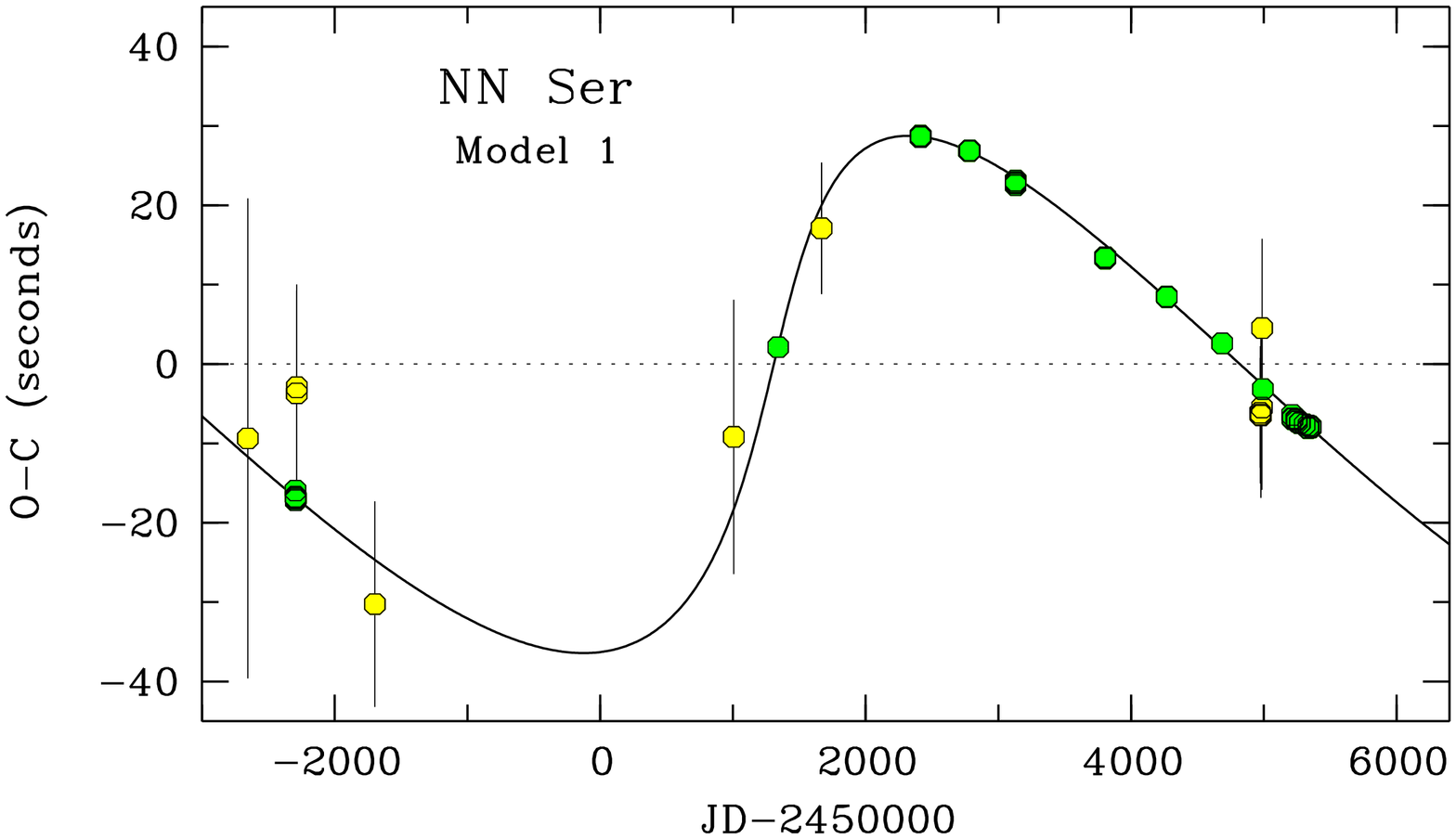}

\vspace*{1mm}
\includegraphics[bb=328 55 540 699,height=89mm,angle=-90,clip]{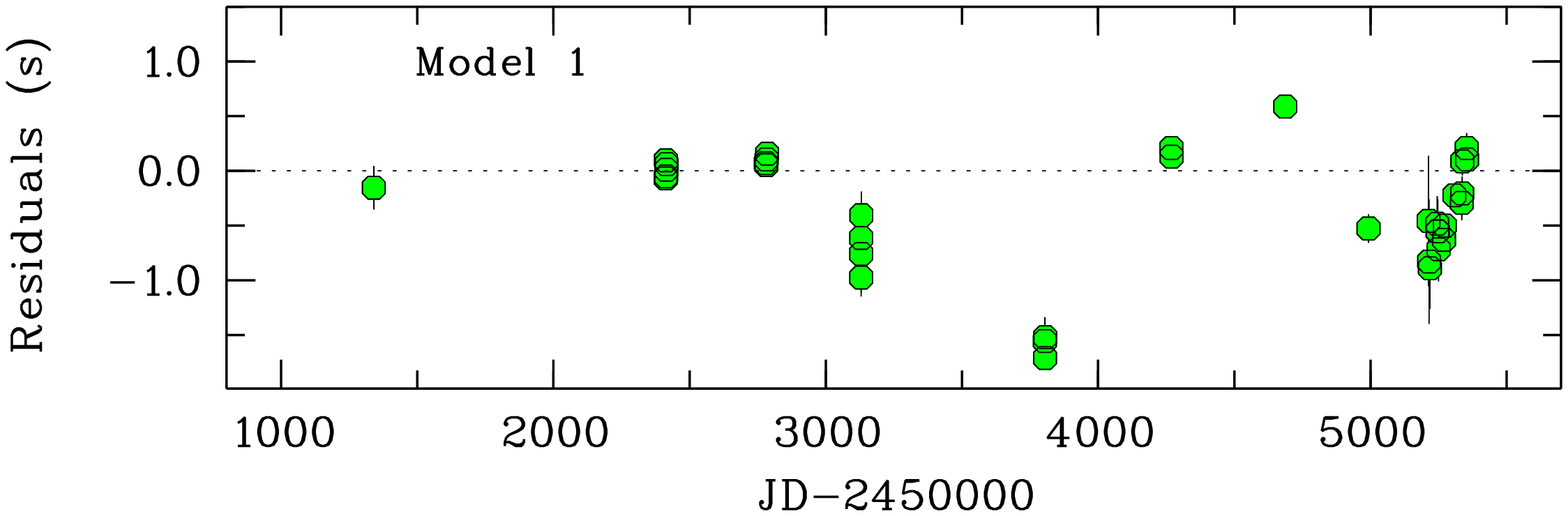}

\vspace*{1mm}
\includegraphics[bb=328 55 540 699,height=89mm,angle=-90,clip]{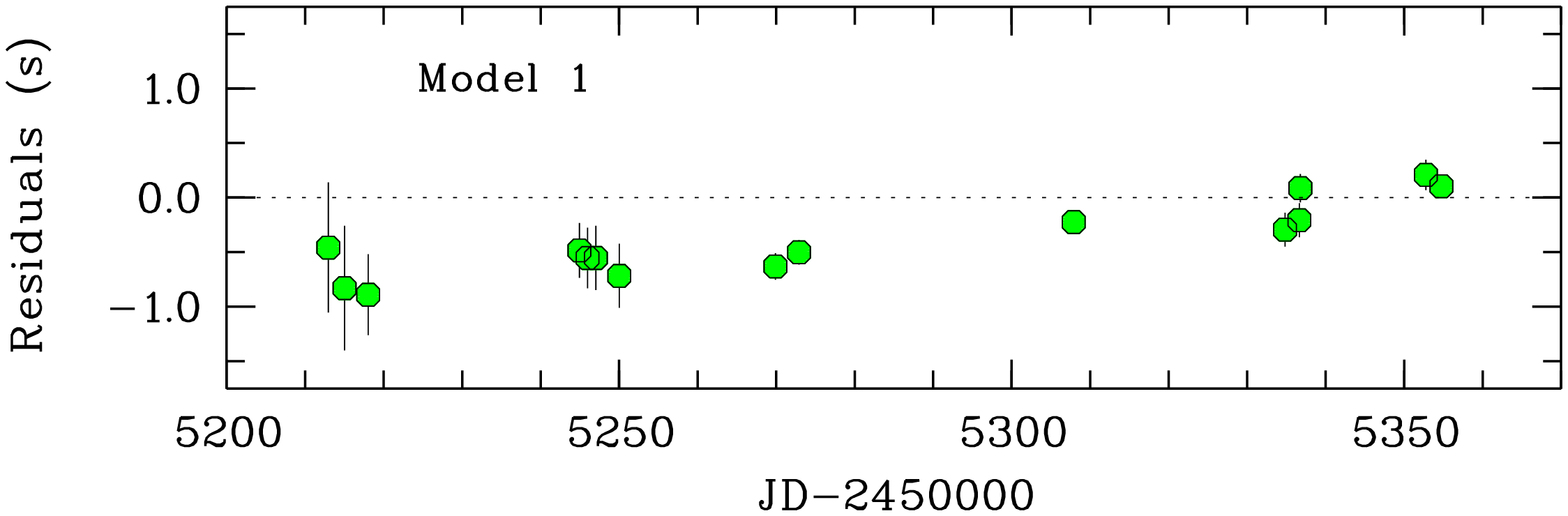}
\caption[chart]{\emph{Top: } \emph{Top: } Observed -- calculated mid
 eclipse time differences relative to the best-fit linear ephemeris
 for Model~1 of a single planet orbiting \nnserab. \emph{Center and
   bottom:} Residuals relative to the eccentric-orbit fit for
 two selected time intervals. }
\vspace{-1mm}
\end{figure}

\begin{figure}[t]
\includegraphics[bb=165 59 540 699,height=89mm,angle=-90,clip]{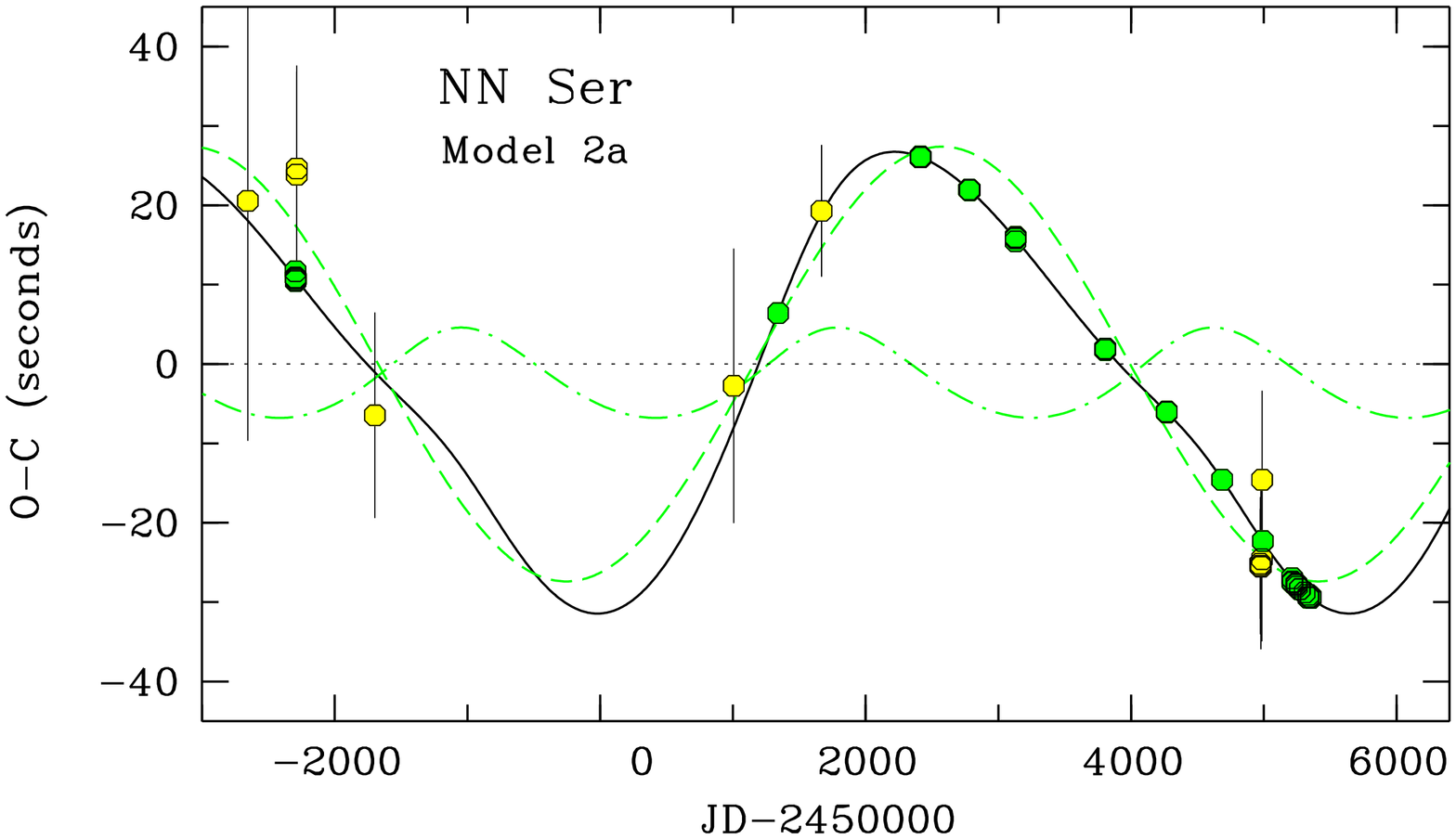}

\vspace*{1mm}
\includegraphics[bb=328 55 540 699,height=89mm,angle=-90,clip]{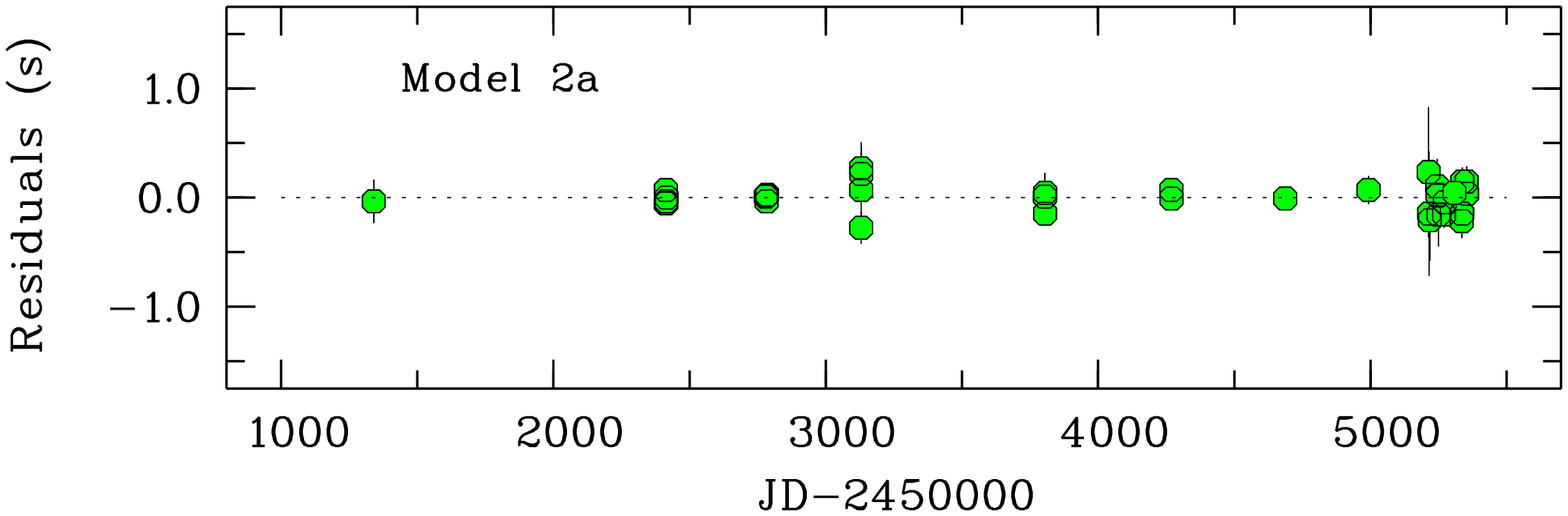}

\vspace*{1mm}
\includegraphics[bb=328 55 540 699,height=89mm,angle=-90,clip]{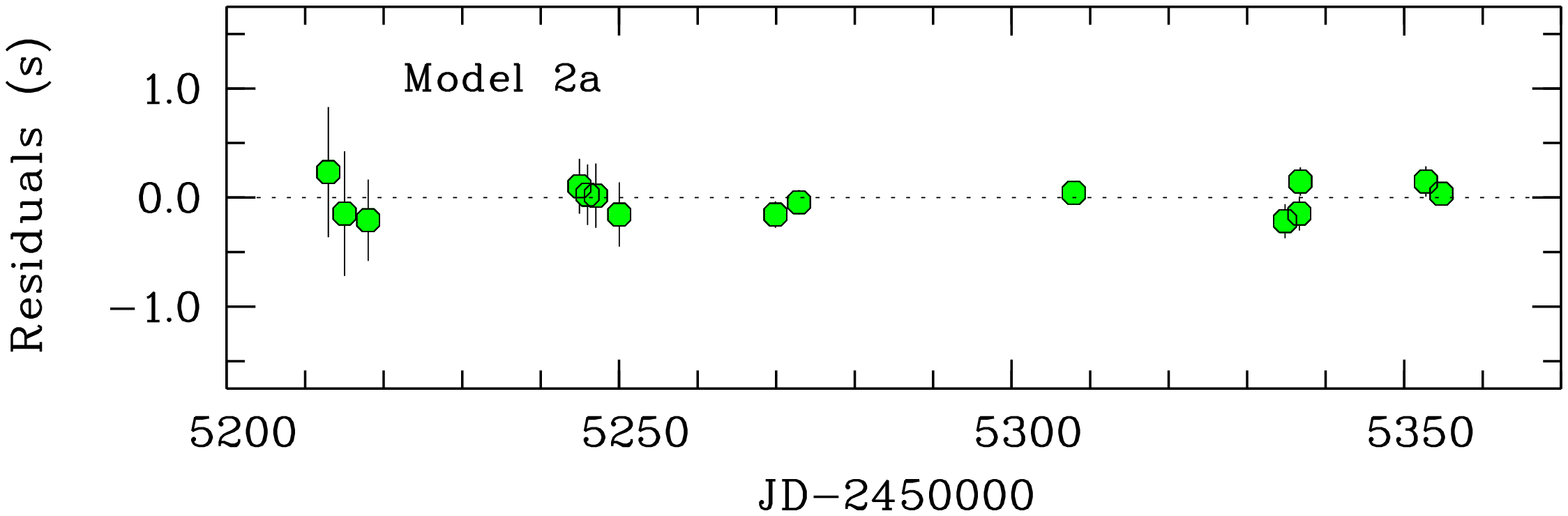}
\caption[chart]{Same as Fig.\,1 for Model~2a of two planets orbiting
 \nnserab. The contributions of components b and c are indicated by
 the dashed and dot-dashed curves, respectively, the solid
 curve shows the combined effect.  }
\vspace{-3mm}
\end{figure}

The mid-eclipse time derived by \citet{haefneretal04} from the trailed
VLT image of 11 June
1999\footnote{http://www.eso.org/public/images/eso9936b/} is the most
variant of the published eclipse time measurements and was assigned a
large error of 17\,s, although this should be a very precise
measurement, given the very simple form of the eclipses in NN\,Ser and
the use of an 8.2m telescope. We reanalysed the image of 11 June 1999,
which started 04:53:05.537 UT with an exposure of 1125.7462\,s and was
taken in good atmospheric conditions. The key issue is the conversion
of the track from pixel space to time. Using two independent methods,
we found that the original analysis by \citet{haefneretal04} was in
error and that the mid-eclipse time can be determined with an accuracy
of 0.20\,s (cycle $E$\,=\,30721). We also reanalysed the less
accurate data of \citet{pigulskimachalska02} (cycle $E$\,=\,33233) by
including the effects of the finite integration times.

Table~2 lists all previously published, the reanalysed, and the new
mid-eclipse times shifted to the solar system barycenter and corrected
for leap seconds. The table also gives the \mbox{1-$\sigma$}
statistical errors and the residuals relative to our Model\,2a, as
shown in Fig.\,2 and discussed in Sect.\,4, below.

\section{The light-travel-time effect in NN Ser}

All measurements of mid-eclipse times of \onnab\ are displayed in
Figs.\,1 and 2 as $O-C$ values relative to the model-dependent linear
ephemerides of the respective fits. Data points with errors $<$1\,s
and $>$1\,s are shown as green and yellow dots, respectively. The
eclipse time measurements dominating the fit are the 1989 data points
of \citet{haefneretal04} near the abscissa value JD'=
\,JD-2450000\,=\,--2295, the reevaluated VLT point on JD'\,=1340, the
2002--2009 series of Warwick eclipse times since JD'\,=\,2411
\citep{parsonsetal10b}, and the data of this work since
JD'\,=\,5212. In particular, the revised VLT mid-eclipse time implies
a twofold change in the time derivative of $O-C$ and excludes the
simple quadratic ephemerides used by \citet{brinkworthetal06} and 
 \citet{parsonsetal10b}. The available data do not exclude abrupt period
changes or an ultimate aperiodicity, but there is no physical process
that predicts such behavior. We consider a periodic behavior the most
promising assumption and proceed to explore this possibility.

Strictly periodic $O-C$ variations may result from apsidal motion of
the binary orbit or an additional body orbiting the binary. Given the
parameters of \nnserab, classical apsidal motion for small
eccentricities $e_\mathrm{bin}$ produces a sinusoidally varying time
shift with an amplitude
$P_\mathrm{bin}e_\mathrm{bin}/\pi$\,=\,3577\,$e_\mathrm{bin}$\,s
\citep{todoran72}. As a result, $e_\mathrm{bin}$\,$\sim$\,0.01 would suffice
to produce the observed amplitude. However, the likewise predicted
variation of the FWHM of the eclipse and the time shift of the
secondary eclipse are not observed
\citep[][this work]{parsonsetal10b}. Furthermore, the observed variation is not
sinusoidal and, given an apsidal motion constant for the secondary
star NN\,Ser\,b of $k_{22}\simeq0.11$, the period of the apsidal
motion would be as short as $\sim$\,0.4 years. Such periodicity is not
detected (see Fig.\,2, bottom panel).

This leaves us with the third-body hypothesis, at least
for the major fraction of the observed eclipse time variations.  In
general, it would be possible that different physical processes
combine to produce the observed signal. We find, however, that a
perfect fit within the very small statistical errors can be obtained
for a signal that consists of the periodicities produced by two
objects orbiting \nnserab. Guided by Ockham's razor and the history of
discoveries in the Solar system, we consider that a fourth body in the
presence of a third one is a natural assumption.

\section{One-planet and two-planet fits to the data}

Including the light-travel-time effect of the objects NN\,Ser\,(ab)c
and NN\,Ser\,(ab)d, the times of mid-eclipse become
\begin{equation}
T=T_0\,+\,P_\mathrm{bin}E + \sum_\mathrm{k=c,d}
\frac{K_\mathrm{bin,k}\,(1-e_\mathrm{k}^2)}{(1+e_\mathrm{k}\,\mathrm{cos}\,\upsilon_\mathrm{k})}\,\mathrm{sin}\,(\upsilon_\mathrm{k}-\varpi_\mathrm{k}),
\end{equation}
where time is measured from a fiducial mid-eclipse time $T_0$.  A
linear binary ephemeris is assumed with $P_\mathrm{bin}$ the orbital
period and $E$ the cycle number. The five free parameters for planet
$k$ are the orbital period $P_\mathrm{k}$, the eccentricity
$e_\mathrm{k}$, the longitude of periastron $\varpi_\mathrm{bin,k}$
measured from the ascending node in the plane of the sky, the time
$T_\mathrm{k}$ of periastron passage, and the amplitude of the eclipse
time variation $K_\mathrm{k}=
a_\mathrm{bin,k}$\,sin\,$i_\mathrm{k}$/c, with $a_\mathrm{bin,k}$ the
semi-major axis of the orbit of the center of mass of the binary about
the common center of mass of the system, $i_\mathrm{k}$ the
inclination, and c the speed of light. In the denominator,
$\upsilon_\mathrm{k}$ is the true anomaly, which progresses through
2$\pi$ over the orbital period $P_\mathrm{k}$.

We explored the multi-dimensional $\,\chi^2$ space of the
two-planet model, using the Levenberg-Marquardt routine implemented in
IDL and an independent code. The search showed that compensation
effects render some parameters ill defined. This uncertainty results,
in particular, from the long hiatus between the accurate measurements
of 1989 \citep{haefneretal04} and 1999 (VLT, this work). We selected the
best model, therefore, by imposing the additional requirement that the
derived orbits be secularly stable.  We investigated all
solutions permitted by the data with numerical N-body simulations with
a variable time step Runge-Kutta integrator, following the orbits over
$10^5$\,yrs, and find that only a narrow range in parameter space
corresponds to stable solutions. In what follows, we consider the
one-planet and the two-planet models in turn. 

\emph{Model\,1} with seven free parameters describes a single planet
with eccentricity $e$. The fit requires $e\!\ga$\,0.60 and is bad for
any value of $e$, with a reduced $\chi^2_\nu\!\ge$\,23.3
(\,$\chi^2$\,=\,1052 for 45 degrees of freedom). The top panel of
Fig\,1 shows the case $e\!=\!0.65$. The residuals based on the
statistical errors of the data points (center panel) reach 23 standard
deviations and indicate that there is an additional modulation at
about half the orbital period. The residuals of the 2010 data (bottom
panel) demonstrate the lack of $O-C$ fluctuations on a short time
scale.

\emph{Model~2} for two planets requires some restriction in
parameters, because the grid search yields good fits for a range of
eccentricities of the outer planet \ec, including zero, and for a
period ratio
$r_\mathrm{p}\!=\!P_\mathrm{c}/P_\mathrm{d}\!=\!1.90\pm0.30$ or
$r_\mathrm{p}\!=\!2.50\pm0.15$ (1-$\sigma$ errors), with the former
slightly preferred. The dichotomy in \rp\ arises from the uncertain
phasing of the singular 1989 point relative to the train of the
1999--2010 data.  Further minima at still larger \rp\ do not
exist. Only a small fraction of the parameter space allowed by the
fits corresponds to secularly stable orbits, however. Near
$r_\mathrm{p}\!\simeq\!2$, orbits with $e_\mathrm{c}\!>\!0.1$ tend to
be unstable, while the stability region is broad in the remaining
parameters for $e_\mathrm{c}\!\approx\!0.02$. Furthermore, all
so\-lu\-tions with $r_\mathrm{p}\!\la\!1.9$ are unstable, with only
some solutions stable at $r_\mathrm{p}\!=\!1.9$. The solutions near
$r_\mathrm{p}\!=\!2.5$ are more generally stable.  We consider
Models~2a and 2b, representing the cases of
$r_\mathrm{p}\!\simeq\!2.0$ and 2.5, respectively, both with
$e_\mathrm{c}\!\equiv\!0$. Model~2a provides the slightly better fit
and is shown in Fig.\,2. It yields $K_\mathrm{c}\!=\!27.4$\,s,
$K_\mathrm{d}\!=\!5.7$\,s, $P_\mathrm{c}\!=\!15.5$\,yrs,
$P_\mathrm{d}\!=\!7.75$\,yrs, and $e_\mathrm{d}\!=\!0.20$ with
$\chi^2_\nu=0.78~(\,\chi^2\!=\!32.9$ for 42 d.o.f.). Periastron
passage of NN\,Ser\,(ab)d \mbox{occurred last on JD'$\!\simeq\!4515$. At
that} time NN\,Ser\,(ab)c was at longitude $213^\circ$. For the low
value of $e_\mathrm{c}\!=\!0.03$, a shallow minimum of \chisq\ is
attained for aligned apses. From the present data, we cannot infer the
true value of \rp\ with certainty, but it is intriguing that objects c
and d may be locked in either the 2\,:\,1 resonance, found also in
other planetary systems, or the 5\,:\,2 resonance. The parameters
for Models~2a and 2b are listed in Table\,1, together with their
1-$\sigma$ errors. A simpler model with two circular orbits reaches
only $\chi^2_\nu=1.96~(\,\chi^2\!=\!86.2$ for 44 d.o.f.) at
$r_\mathrm{p}\!=\!2.46$ and can be excluded.

\begin{table*}[t]
\begin{flushleft}
\caption{Parameters of the models fitted to the measured mid-eclipse
 time variations of NN\,Ser, where $\equiv$ indicates a fixed
 parameter.}
\begin{tabular}{c@{\hspace{2mm}}c@{\hspace{3mm}}cccccccccccc@{\hspace{3mm}}c@{\hspace{5mm}}c}\\[-1ex]
\hline \hline\\[-1ex]
Model & Planets &Number    & \pc\  & \pd\ &\pc/\pd\ & \ec\ & \ed\ & \ac\ & \ad\ & \omegac\ & \omegad\ & \massc\,sin\,$i_\mathrm{c}$ & \massd\,sin\,$i_\mathrm{d}$ & \chisq &\chisqr\ \\ 
     &      & free par. & (yrs) & (yrs)&         &      &      & (AU) & (AU) & ($^\circ$)&($^\circ$)& (\mjup)&(\mjup)& & \\  [1ex]
\hline \\ [-1ex]
\hspace{-1.4mm}1     & 1    & 2\,+\,5  & 22.60 &      &         & \hspace{-1.0mm}$\ga\!0.65$&& 6.91 &      &  8.0     &          &  8.36 &      & \hspace{-1.9mm}1052.3 & \hspace{-0.8mm}23.38  \\[0.5ex]
2a     & 2    & 2\,+\,8 & 15.50 & 7.75 & $2.00$  &\hspace{-1.0mm}$\equiv 0.0$ &0.20    & 5.38 & 3.39 & &  74    &  6.91 & 2.28 & \hspace{1.4mm}32.9& 0.78\\
     &      &          &$\hspace{-0.8mm}\pm 0.45$ & $\hspace{-1.8mm}\pm 0.35$ &$\hspace{-1.8mm}\pm 0.15$& &$\hspace{-1.8mm}\pm 0.02$&$\hspace{-1.8mm}\pm 0.20$&$\hspace{-1.5mm}\pm 0.10$& &$\hspace{-0.5mm}\pm 4$&$\hspace{-2.0mm}\pm 0.54$&$\hspace{-1.8mm}\pm 0.38$& & \\[0.5ex]
2b     & 2    & 2\,+\,8  & 16.73 & 6.69 & 2.50 &\hspace{-1.0mm}$\equiv 0.0$ & 0.23  & 5.66 & 3.07 &     &  73   &  5.92 & 1.60 & \hspace{1.4mm}33.8& 0.80\\
     &      &          &$\hspace{-0.8mm}\pm 0.26$ & $\hspace{-1.8mm}\pm 0.40$ &$\hspace{-1.8mm}\pm 0.15$& &$\hspace{-1.8mm}\pm 0.04$&$\hspace{-2.2mm}\pm 0.06$&$\hspace{-1.5mm}\pm 0.13$& &$\hspace{-0.5mm}\pm 7$&$\hspace{-2.0mm}\pm 0.40$&$\hspace{-1.8mm}\pm 0.27$& & \\[1ex]
\hline\\[-3ex]
\end{tabular}
\end{flushleft}
\end{table*}

Using Model\,2a as input to our N-body simulations, we find that
\ec\ and \ed\ oscillate around 0.02 and 0.22 with amplitudes of 0.02
and 0.05, respectively. The difference $\Delta \varpi$ of the
periastron longitudes circulates on a time scale of 400\,yrs.  The
periods perform small-amplitude anti-phased oscillations, which cause
\rp\ to oscillate between 1.9 and 2.2. Even if the two planets are
secularly locked in the 2\,:\,1 mean motion resonance, therefore, the
observed period ratio at any given time may deviate slightly from its
nominal value.

For Model~2a, the best-fit binary ephemeris is $T$\,= BJED
2,447344.524425(40)\,+\,0.1300801419(10)\,$E$, where the errors refer
to the last digits. Adding a quadratic term $BE^2$ to the ephemeris
does not improve the two-planet fit and yields a 1-$\sigma$ limit of
$\mid\!B\!\mid\,<\,1.5\,10^{-13}$ days, leaving room for a period
change by gravitational radiation or a long-term activity-related
effect \citep{brinkworthetal06,parsonsetal10b}.

\section{Discussion}

The large amplitude of the $O-C$ eclipse time variations in
\nnser\ can only be explained by a third body in the system, while the
still substantial residuals from a single-planet fit could, in
principle, have a different origin from that of a fourth body. The
two-planet model, however, possesses the beauty of simplicity, and the
fact that the residuals for the entire data set vanish simultaneously
imposes tight restrictions on any other mechanism. In particular, the
lack of short-term variability of the residuals in the first half of
2010 argues against any process that acts on a short time scale or
leads to erratic eclipse time variations. Hence, there is strong
evidence for two planets orbiting \nnserab.

With masses
$M_\mathrm{c}\mathrm{sin}\,i_\mathrm{c}\!\simeq\!6$\,\mjup\ and
$M_\mathrm{d}\mathrm{sin}\,i_\mathrm{d}\!\simeq\!2$\,\mjup, NN\,Ser\,(ab)c
and NN\,Ser\,(ab)d both qualify as giant planets for all inclinations
$i_\mathrm{\,c}>$\,28$^\circ$ and $i_\mathrm{\,d}>$\,9$^\circ$,
respectively. The probable detection of resonant motion with a period
ratio of either 2\,:\,1 or 5\,:\,2 is a major bonus, which adds to the
credence of the two-planet model. It is the second planetary
 system found by eclipse timing, after HW\,Vir \citep{leeetal09}.

Given a pair of planets orbiting a post-CE binary, two formation
scenarios are possible. They could either be old first-generation
planets that formed in a circumbinary protoplanetary disk or they
could be young second-generation planets formed $\la\!10^6$ yrs ago in
a disk that resulted from the CE \citep{perets10}. To
evaluate both scenarios, we have reconstructed the CE evolution of
NN\,Ser\,ab using the improved algorithm by \citet{zorotovicetal10},
who constrain the CE efficiency to a range $\alpha\simeq0.2 - 0.3$.
Possible solutions for the progenitor binary of NN\,Ser\,ab are
not very sensitive to $\alpha$: for $\alpha=0.25$,
the progenitor was a giant of 2.08\,\msun\ and  radius
194\,\rsun\ with the present secondary star at a separation of
1.44\,AU. When the CE engulfed the secondary star, dynamic friction
caused the latter to spiral in rapidly, thereby dramatically decreasing the
binary separation to the current
0.0043\,AU.
Stability arguments imply that any planet from the pre-CE phase must
have formed with semi-major axes exceeding 3.5\,AU
\citep{holmanwiegert99}. With three quarters of the central mass
expelled in the CE event, pre-existing planets would move outward or
may even be lost from the system. However, given a sufficiently dense
and slowly expanding CE, the dynamical force experienced by them may
have ultimately moved them inward \citep{alexanderetal76}. Since the
drag primarily affects the more massive and more slowly moving outer
planet, such a scenario could lead to resonant orbits, so a
first-generation origin appears possible.

The alternative post-CE origin in a second-generation of planet
formation is also possible, since the formation of circumbinary disks
is a common phenomenon among post-AGB binary stars and the
concentration of a slow, dusty wind to the orbital plane of the binary
is thought to favor the formation of planets
\citep[e.g.][]{vanwinckeletal09,perets10}. In particular the tiny
separation of the present binary poses no problem for stable orbits of
second-generation planets even at significantly shorter distances than
the inner planet that we have detected \citep{holmanwiegert99}. A
particularly intriguing aspect of a second-generation origin of the
planets in NN~Ser would be their extreme youth, equal to or less than
the $10^6$\,yrs cooling age of the white dwarf \citep{wood95}. This
feature would distinguish them from all known exoplanets and may
ultimately lead to their direct detection. While we cannot presently
prove a second-generation origin for these planets, modeling
the CE event may allow us to distinguish between the two scenarios.


\begin{acknowledgements}
We would like to thank Dr. Reinhold Haefner for information concerning
the original VLT observations and analyses and Dr. Andrzej Pigulski
for sending us his original photometry.  This work is based on data
obtained with the MONET telescopes funded by the "Astronomie \&
Internet" program of the Alfried Krupp von Bohlen und Halbach
Foundation, Essen, on observations with the ESO NTT under ESO
programme 085.D-0541, and on data obtained from the ESO/ST-ECF Science
Archive Facility. TRM, VSD, CMC, and SPL acknowledge grant support from
the UK's STFC. MRS acknowledges support from FONDECYT under grant
number 1061199 (MRS) and the Centre of Astrophysics
Valpara{\'i}so. DEW acknowledges the support of the Norman Hackerman
Advanced Research Program under grant 003658-0255-2007.
\end{acknowledgements}

\bibliographystyle{aa}



\begin{table*}[t]
\begin{flushleft}
\caption{Previously published, reanalysed, and new mid-eclipse times of the white dwarf in
 NN\,Ser with residuals for the light-travel-time effect produced
 by the two planets of Model 2a. The published mid-eclipse times have been converted to BJD(TT) if not yet on this time standard.}
\begin{tabular}{r@{\hspace{8mm}}c@{\hspace{8mm}}c@{\hspace{8mm}}c@{\hspace{6mm}}r@{\hspace{6mm}}r@{\hspace{6mm}}c@{\hspace{6mm}}l}
\hline \hline\\[-1.5ex]
E & BJD(TT) & Error & Residual  & Error & Residual & References & Comment  \\
& JD2400000$+$&(days) & (days) & (s)  & (s) \\[0.5ex]
\hline\\
    0 &  47344.5246635 &    0.0003500  &  0.0000290 & 30.00 &  2.51 & (1) & Reanalysed\\
 2760 &  47703.5457436 &    0.0000020  &  0.0000012 &  0.17 &  0.10 & (2) & \\
 2761 &  47703.6758326 &    0.0000060  &  0.0000101 &  0.52 &  0.87 & (2) & \\
 2769 &  47704.7164596 &    0.0000030  & \hspace{-2.0mm}$-$0.0000038 &  0.26 & \hspace{-2.0mm}$-$0.33 & (3) & \\
 2776 &  47705.6270226 &    0.0000030  & \hspace{-2.0mm}$-$0.0000016 &  0.26 & \hspace{-2.0mm}$-$0.14 & (3) & \\
 2777 &  47705.7571046 &    0.0000070  &  0.0000003 &  0.60 &  0.03 & (3) &  Corrected\\
 2831 &  47712.7815836 &    0.0001500  &  0.0001534 & 12.96 & 13.25 & (2) & \\
 2839 &  47713.8222336 &    0.0001500  &  0.0001625 & 12.96 & 14.04 & (2) & \\
 7360 &  48301.9141954 &    0.0001500  & \hspace{-2.0mm}$-$0.0000627 & 12.96 & -5.42 & (2) & \\
28152 &  51006.5405495 &    0.0002000  &  0.0000605 & 17.28 &  5.23 & (2) & \\
30721 &  51340.7165402 &    0.0000023  & \hspace{-2.0mm}$-$0.0000004 &  0.20 & \hspace{-2.0mm}$-$0.03 & (2) & Reanalysed\\
33233 &  51667.4780058 &    0.0000960  &  0.0000041 &  8.29 &  0.35 & (4) & Reanalysed\\
38960 &  52412.4470566 &    0.0000006  & \hspace{-2.0mm}$-$0.0000006 &  0.05 & \hspace{-2.0mm}$-$0.05 & (5) & \\
38961 &  52412.5771382 &    0.0000005  &  0.0000008 &  0.04 &  0.07 & (5) & \\
38968 &  52413.4876977 &    0.0000009  & \hspace{-2.0mm}$-$0.0000006 &  0.08 & \hspace{-2.0mm}$-$0.05 & (5) & Corrected\\
38976 &  52414.5283389 &    0.0000007  & \hspace{-2.0mm}$-$0.0000004 &  0.06 & \hspace{-2.0mm}$-$0.03 & (5) & \\
38984 &  52415.5689804 &    0.0000007  &  0.0000000 &  0.06 &  0.00 & (5) & \\
41782 &  52779.5331703 &    0.0000015  &  0.0000001 &  0.13 &  0.01 & (5) & \\
41798 &  52781.6144523 &    0.0000007  &  0.0000002 &  0.06 &  0.02 & (5) & \\
41806 &  52782.6550927 &    0.0000008  & \hspace{-2.0mm}$-$0.0000004 &  0.07 & \hspace{-2.0mm}$-$0.03 & (5) & \\
41820 &  52784.4762150 &    0.0000008  &  0.0000003 &  0.07 &  0.03 & (5) & \\
44472 &  53129.4486808 &    0.0000040  &  0.0000008 &  0.35 &  0.07 & (5) & \\
44473 &  53129.5787632 &    0.0000028  &  0.0000031 &  0.24 &  0.27 & (5) & \\
44474 &  53129.7088370 &    0.0000017  & \hspace{-2.0mm}$-$0.0000032 &  0.15 & \hspace{-2.0mm}$-$0.28 & (5) & \\
44480 &  53130.4893234 &    0.0000030  &  0.0000025 &  0.26 &  0.22 & (5) & \\
49662 &  53804.5644567 &    0.0000025  &  0.0000001 &  0.22 &  0.01 & (5) & \\
49663 &  53804.6945350 &    0.0000012  & \hspace{-2.0mm}$-$0.0000017 &  0.10 & \hspace{-2.0mm}$-$0.15 & (5) & \\
49671 &  53805.7351781 &    0.0000006  &  0.0000005 &  0.05 &  0.04 & (5) & \\
53230 &  54268.6903114 &    0.0000006  &  0.0000008 &  0.05 &  0.07 & (5) & \\
53237 &  54269.6008713 &    0.0000002  & \hspace{-2.0mm}$-$0.0000001 &  0.02 & \hspace{-2.0mm}$-$0.01 & (5) & \\
56442 &  54686.5076279 &    0.0000009  & \hspace{-2.0mm}$-$0.0000001 &  0.08 & \hspace{-2.0mm}$-$0.01 & (5) & \\
58638 &  54972.1634971 &    0.0000800  & \hspace{-2.0mm}$-$0.0000380 &  6.91 & -3.28 & (6) & \\
58645 &  54973.0740553 &    0.0001000  & \hspace{-2.0mm}$-$0.0000406 &  8.64 & -3.51 & (6) & \\
58684 &  54978.1471791 &    0.0001200  & \hspace{-2.0mm}$-$0.0000408 & 10.37 & -3.53 & (6) & \\
58745 &  54986.0820789 &    0.0001200  & \hspace{-2.0mm}$-$0.0000274 & 10.37 & -2.37 & (6) & \\
58753 &  54987.1228359 &    0.0001300  &  0.0000887 & 11.23 &  7.66 & (6) & \\
58796 &  54992.7161925 &    0.0000015  &  0.0000008 &  0.13 &  0.07 & (6) & \\
60489 &  55212.9418187 &    0.0000069  &  0.0000027 &  0.60 &  0.23 & (7,8) & \\
60505 &  55215.0230961 &    0.0000066  & \hspace{-2.0mm}$-$0.0000017 &  0.57 & \hspace{-2.0mm}$-$0.15 & (7,8) & \\
60528 &  55218.0149380 &    0.0000043  & \hspace{-2.0mm}$-$0.0000024 &  0.37 & \hspace{-2.0mm}$-$0.21 & (7,8) & \\
60735 &  55244.9415254 &    0.0000029  &  0.0000012 &  0.25 &  0.10 & (7,8) & \\
60743 &  55245.9821654 &    0.0000032  &  0.0000003 &  0.28 &  0.03 & (7,8) & \\
60751 &  55247.0228063 &    0.0000034  &  0.0000002 &  0.29 &  0.02 & (7,8) & \\
60774 &  55250.0146469 &    0.0000034  & \hspace{-2.0mm}$-$0.0000018 &  0.29 & \hspace{-2.0mm}$-$0.16 & (7,8) & \\
60927 &  55269.9169047 &    0.0000014  & \hspace{-2.0mm}$-$0.0000018 &  0.12 & \hspace{-2.0mm}$-$0.16 & (7,9) & \\
60950 &  55272.9087487 &    0.0000013  & \hspace{-2.0mm}$-$0.0000005 &  0.11 & \hspace{-2.0mm}$-$0.04 & (7,9) & \\
61219 &  55307.9003015 &    0.0000010  &  0.0000005 &  0.09 &  0.04 & (7,10) & \\
61426 &  55334.8268834 &    0.0000018  & \hspace{-2.0mm}$-$0.0000025 &  0.16 & \hspace{-2.0mm}$-$0.22 & (7,9) & \\
61440 &  55336.6480059 &    0.0000018  & \hspace{-2.0mm}$-$0.0000017 &  0.16 & \hspace{-2.0mm}$-$0.15 & (7,9) & \\
61441 &  55336.7780894 &    0.0000015  &  0.0000017 &  0.13 &  0.15 & (7,9) & \\
61564 &  55352.7779443 &    0.0000016  &  0.0000017 &  0.14 &  0.15 & (7,9) & \\
61579 &  55354.7291448 &    0.0000009  &  0.0000004 &  0.08 &  0.03 & (7,10) & \\ [1.5ex]
\hline\\[-2ex]
\end{tabular}\\

(1) Haefner, R., 1989, ESO Msngr, 55, 61, reanalysed using up-to-date
eclipse profile; (2) Haefner et al. (2004), misprint for E=2777
corrected, VLT trailed imaging observation (E=30721) reanalysed using
the original data; (3) Wood, J.~H. \& Marsh, T.~R., 1991, ApJ, 381,
551; (4) Pigulski \& Michalska (2002), reanalysed using the original
data; (5) Parsons et al. (2010b), timing for E=38968 corrected for
misprint; (6) Qian et al. (2009); (7) This work, (8) MONET/North 1.2-m
white light photometry, (9) McDonald 2.1-m photometry with Schott BG40
filter, (10) ESO NTT 3.5-m ULTRACAM Sloan g' photometry.
\end{flushleft}
\end{table*}

\end{document}